\documentclass[10pt, conference, letterpaper]{IEEEtran}
\usepackage{hyperref}

\IEEEoverridecommandlockouts
\usepackage{cite}
\usepackage{amsmath,amssymb,amsfonts}

\usepackage{algorithm}
\usepackage{algpseudocode}

\usepackage{graphicx}
\usepackage{newtxmath}
\usepackage{lipsum}
\usepackage{subfigure}
\usepackage{textcomp}
\usepackage{xcolor}
\usepackage{bm}
\usepackage{url}
\usepackage{tabularx,booktabs}

\usepackage{tabularx,booktabs}

\usepackage{tabularray}
\UseTblrLibrary{booktabs}

\newif\ifcolormarker
\colormarkertrue

\algnewcommand\algorithmicforeach{\textbf{for each}}
\algdef{S}[FOR]{ForEach}[1]{\algorithmicforeach\ #1\ \algorithmicdo}

\def\BibTeX{{\rm B\kern-.05em{\sc i\kern-.025em b}\kern-.08em
    T\kern-.1667em\lower.7ex\hbox{E}\kern-.125emX}}
\begin{document}

\title{Leveraging Spatial and Temporal Correlations for Network Traffic Compression}


\author{\IEEEauthorblockN{ Paul~Almasan\IEEEauthorrefmark{1}, Krzysztof~Rusek\IEEEauthorrefmark{2},
Shihan Xiao\IEEEauthorrefmark{3},
Xiang Shi\IEEEauthorrefmark{3},
Xiangle Cheng\IEEEauthorrefmark{3},\\
Albert~Cabellos-Aparicio\IEEEauthorrefmark{1},
Pere~Barlet-Ros\IEEEauthorrefmark{1}}
\IEEEauthorblockA{\IEEEauthorrefmark{1}Barcelona Neural Networking Center, 
Universitat Politècnica de Catalunya}
\IEEEauthorblockA{\IEEEauthorrefmark{2}AGH University of Science and Technology, Institute of Telecommunications\\
}
\IEEEauthorblockA{\IEEEauthorrefmark{3}Huawei Technologies}
}

\maketitle

\begin{abstract}
The deployment of modern network applications is increasing the network size and traffic volumes at an unprecedented pace. Storing network-related information (e.g., traffic traces) is key to enable efficient network management. However, this task is becoming more challenging due to the ever-increasing data transmission rates and traffic volumes. In this paper, we present a novel method for network traffic compression that exploits spatial and temporal patterns naturally present in network traffic. We consider a realistic scenario where traffic measurements are performed at multiple links of a network topology using tools like SNMP or NetFlow. Such measurements can be seen as multiple time series that exhibit spatial and temporal correlations induced by the network topology, routing or user behavior. Our method leverages graph learning methods to effectively exploit both types of correlations for traffic compression. The experimental results show that our solution is able to outperform GZIP, the \textit{de facto} traffic compression method, improving by 50\%-65\% the compression ratio on three real-world networks.
\end{abstract}


\section{Introduction}

In the last years, modern networks have seen
a considerable growth in network traffic\cite{tune2013internet} and connected devices \cite{alam2018reliable}. The deployment of modern applications (e.g., vehicular networks, IoT, virtual reality, video streaming, Industry 4.0) and continuous improvements in network technology (e.g., link speeds) are accentuating this trend even further. Storing network traffic information (e.g., packet traces, link-level traffic measurements, flow-level measurements) is important for network operators to perform network management tasks, such as network planning, traffic engineering, traffic classification, anomaly detection or network forensics, among others. The emerging Network Digital Twin paradigm (NDT) will also require storage and analysis of vast amounts of network traffic data~\cite{9374645, 9795043}.

As a consequence, the efficient storage of network traffic is becoming more challenging than ever. Network traffic traces from Internet Service Providers (ISP), backbone or data center networks can easily occupy hundreds of terabytes per day~\cite{10.1145/2829988.2787472} or petabytes in the case of mobile networks~\cite{7762185}. For example, only a 24-hour trace from a single 10 Gbps link can result in 108 terabytes of data in the worst case. Storing traces from real-world networks can be difficult as they have in the order of hundreds of links~\cite{knight2011internet}. In addition, such traces can contain thousands of concurrent flows per second~\cite{10.1145/2829988.2787472}. Even storing aggregated flow-level information (e.g., NetFlow) can require hundreds of terabytes of disk storage per day\cite{fang1999inter, mori2004characteristics}. 

Traditionally, network traffic traces are compressed using GZIP~\cite{gzip_reference}, a popular lossless method for compressing files regardless of their format (e.g., text, csv file). Network operators typically collect traffic traces in PCAP format \cite{pacpfile} and they simply compress them with GZIP or similar tools. However, GZIP is a generic compression tool, resulting in sub-optimal compression performance when applied to network traffic data.

Past works showed that network traffic traces are far from being purely random, meaning that they intrinsically have some underlying structure~\cite{tune2013internet, 10.1145/1879141.1879175, lan2006measurement, gao2020predicting, gao2020incorporating}. In particular, traffic traces are known to present spatial and temporal patterns that could potentially be exploited to increase current compression ratios. 
In this work, we seek to understand if recent advancements in neural network (NN) architectures could effectively be used to leverage such correlations to achieve better compression ratios than traditional tools such as GZIP.

We consider a traffic compression scenario where multiple link-level traffic measurements are performed over time for a network topology using standard measurement tools, such as SNMP or NetFlow. These measurements can be seen as multiple time series (i.e., one per link) that exhibit spatio-temporal correlations.
Temporal correlations result from user behavior and seasonality in network traffic (e.g., day/night or workday/weekend patterns). Spatial correlations between links are mostly induced by the network topology and routing, among other reasons (e.g., correlations in traffic demands or resulting from protocol behavior).

In this paper we present a neural traffic compression method that exploits the spatio-temporal correlations naturally present in network traffic. Our compressor contains two main modules: a \textit{predictor} that is implemented using neural networks (i.e., Recurrent and Spatio-Temporal Graph Neural Networks) and an \textit{encoder}. The main role of the predictor is to exploit the spatial and temporal correlations between the network links to accurately estimate, from past observations, the distribution of the data to be compressed. The encoder is implemented using arithmetic coding (AC)~\cite{10.1145/214762.214771}, a popular lossless compression method. Based on the predicted distributions, AC decides how to better encode the traffic information. The proposed solution also implements a \textit{decoder} for decompression, which inverts the process to recover the original traffic data.

To showcase the compression capabilities of our method, we first evaluate it on synthetically-generated traffic with different degrees of temporal and spatial correlation. The results with synthetic data show that our proposed solution can improve GZIP's compression ratios by $\geq$35\%, even in scenarios with weak correlation. Next, we evaluate our compression method with real-world datasets that cover several months of traffic from three real-world networks. Experimental results show that our method can reduce the size of compressed files by 50\%-65\% compared to GZIP and by a factor between 2.6x and 4.2x with respect to the original file.

\section{Background}

\begin{figure}[!t]
  \centering
  \includegraphics[width=0.8\linewidth]{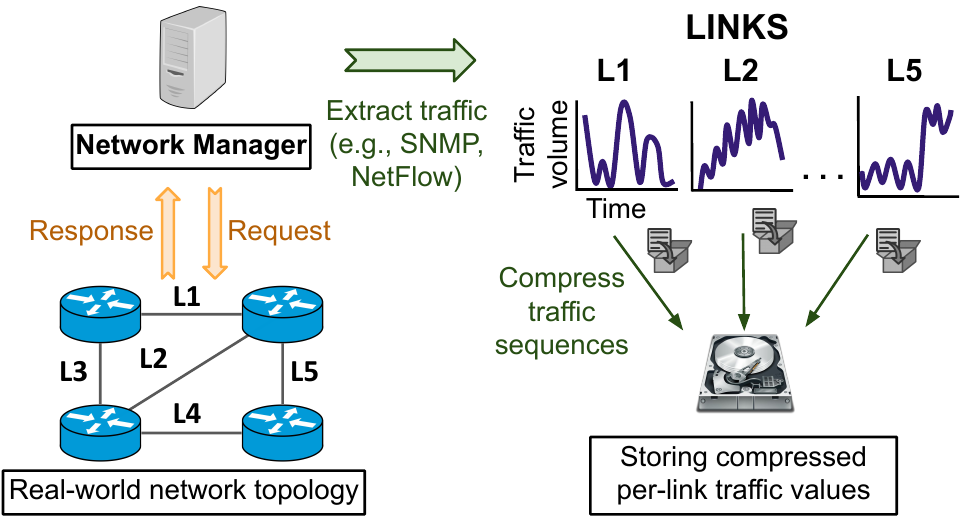}
  \caption{Overview of the network traffic compression scenario. Link-level traffic measurements are extracted from the real-world network and stored in the disk.}
  \label{fig:compress_scenario}
\vspace{-0.4cm}
\end{figure}

In this paper, we consider the compression scenario defined by a network topology with link-level traffic measurements. These measurements indicate the traffic volume over time going through each link. They can be obtained from the real-world network using network monitoring tools such as SNMP or NetFlow. Link-level measurements are performed periodically and stored in time bins (e.g., bins of 5 minutes), resulting in a sequence of accumulated traffic values that we want to store efficiently in disk\footnote{We chose this scenario for its relevance and simplicity, but note that the same principles apply for example to flow-level measurements (e.g., NetFlow), where flows (instead of link-level measurements) can be seen as multiple time series exhibiting spatio-temporal patterns.}. Figure~\ref{fig:compress_scenario} shows an overview of the compression scenario.

When compressing network traffic measurements, network administrators  typically follow a simplistic approach based on well-known compression software such as GZIP~\cite{gzip_reference}. However, such methods are generic, meaning that they were designed to compress multiple kinds of information (e.g., images, text). This results in low compression ratios when used with network traffic traces. Equation~\ref{equation:comp_ratio} shows how to compute the compression ratio (CR) of a file.

\begin{equation}
CR = \frac{Uncompressed\_size}{Compressed\_size}
\label{equation:comp_ratio}
\end{equation}

\subsection{Exploiting temporal and spatial correlations}
\label{subsec:exploiting_temp_spati}

\begin{figure}[!t]
  \centering
  \includegraphics[width=0.9\columnwidth]{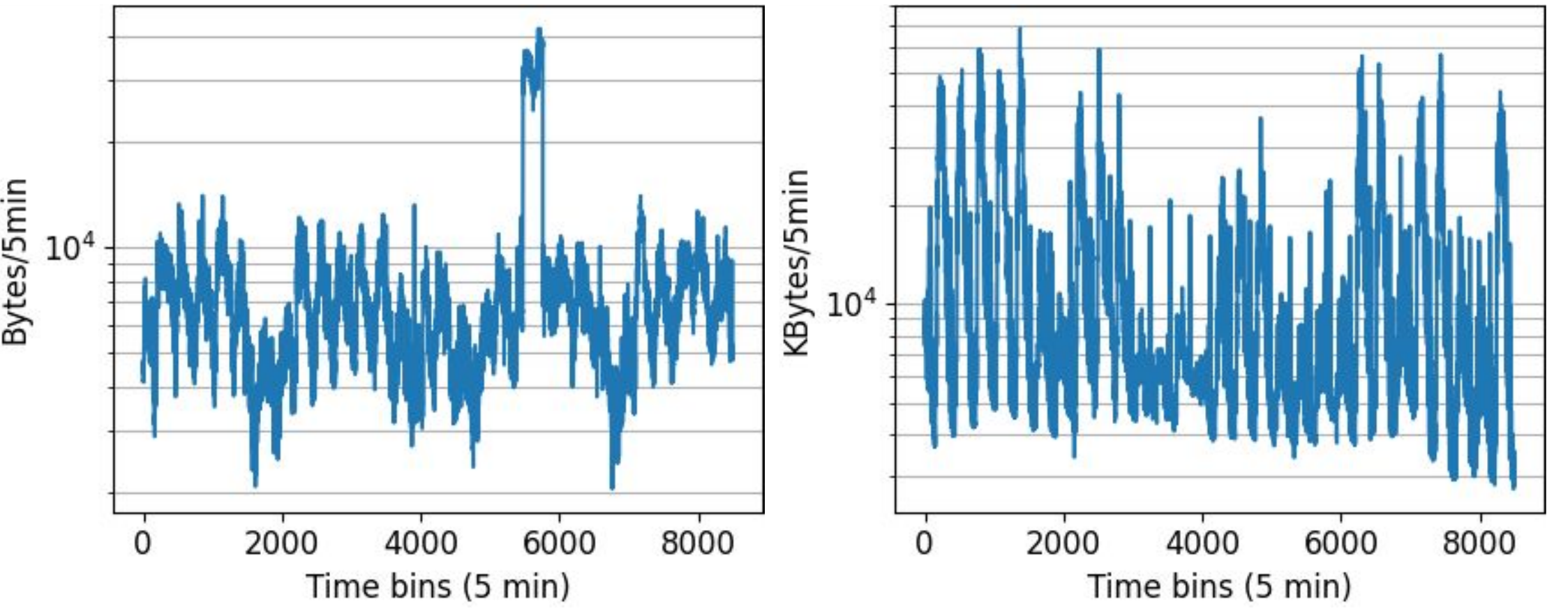}
  \caption{Two link-level network traffic measurements during $\approx$1 month from two real-world datasets. Notice the y-axis is in logarithmic scale. The figures indicate that the resulting time series from the measurements have temporal patterns.}
  \label{fig:temporal_correlation}
 \vspace{-0.2cm}
\end{figure}

In our work, we leverage ML to exploit the network traffic characteristics and achieve high compression ratios. Specifically, link-level traffic measurements are time series data, meaning that the traffic values can be seen as a set indexed by time. A time series can be typically described by its seasonality and trend. Seasonality refers to a pattern repeated in time at a certain frequency (e.g., day/night). The trend indicates long-term tendency of the time series to increase, decrease or remain stable. Figure~\ref{fig:temporal_correlation} shows the daily seasonality present in link-level traffic measurements during $\approx$1 month on two real-world datasets used in our experiments (see Section~\ref{subsec:methodology}). In addition, the network topology and routing introduce spatial correlations in the link-level traffic measurements. This means that the links sharing paths are going to have a similar traffic behavior, which we believe that it can be exploited for improving the compression ratios.

\begin{figure}[!b]
  \centering
  \includegraphics[width=0.9\linewidth]{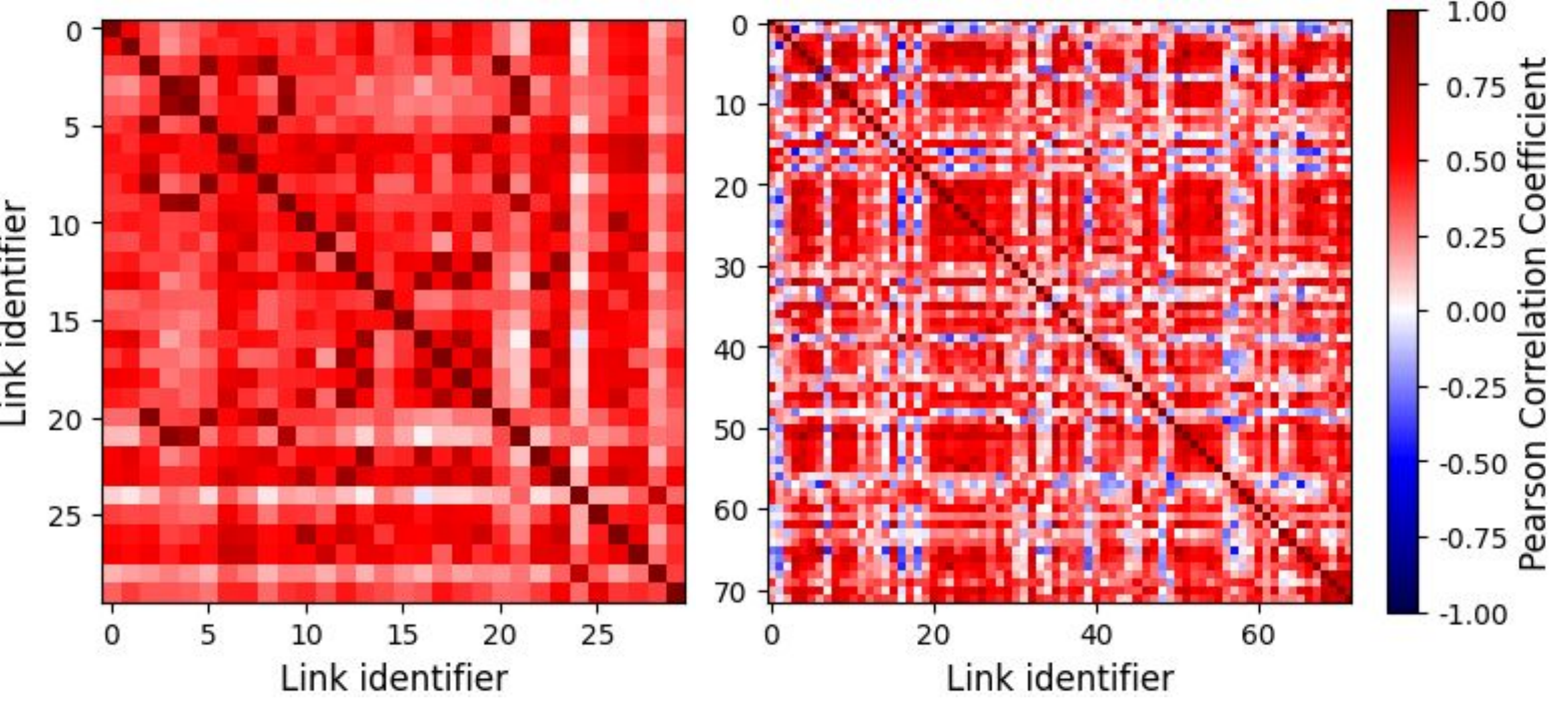}
  \caption{Pearson correlation between links for the Abilene (left) and Geant (right) datasets. The darker colors indicate high spatial correlation between links. This means that the traffic values between links have a positive correlation (red) or a negative one (blue). Best viewed in color.}
  \label{fig:pearson_correlation}
\end{figure}

To showcase the presence of spatial correlations, we compute the Pearson correlation coefficient between each pair of links in two real-world topologies. This coefficient indicates how strongly correlated are two sets of data or vectors. The following equation shows how to compute the Pearson correlation:

\begin{equation}
r = \frac{{}\sum_{i=1}^{n} (x_i - \overline{x})(y_i - \overline{y})}
{\sqrt{\sum_{i=1}^{n} (x_i - \overline{x})^2(y_i - \overline{y})^2}}
\end{equation}
\label{equation:pearson_correlation}

where $\overline{x}$ and $\overline{y}$ are the means of the vectors x and y respectively. The resulting value is contained between the range [-1, +1], where -1 indicates negative correlation. This can happen when the traffic increases in one link but decreases in the other link. A value of 0 indicates no correlation and values close to +1 indicate positive correlation (i.e., the traffic increases in both links in similar proportions). Figure~\ref{fig:pearson_correlation} shows the Pearson correlation for the real-world Abilene (left) and Geant (right) datasets~\cite{SNDlib10}. The darker the color in the figure, the higher is the correlation between links. The figures indicate that indeed there is spatial correlation between links. We believe that both temporal and spatial correlations can be exploited to achieve higher compression ratios than generic methods like GZIP.

\subsection{Arithmetic Coding}
\label{subsec:arithm_coding}

\begin{figure}[!t]
  \centering
  \includegraphics[width=0.7\linewidth]{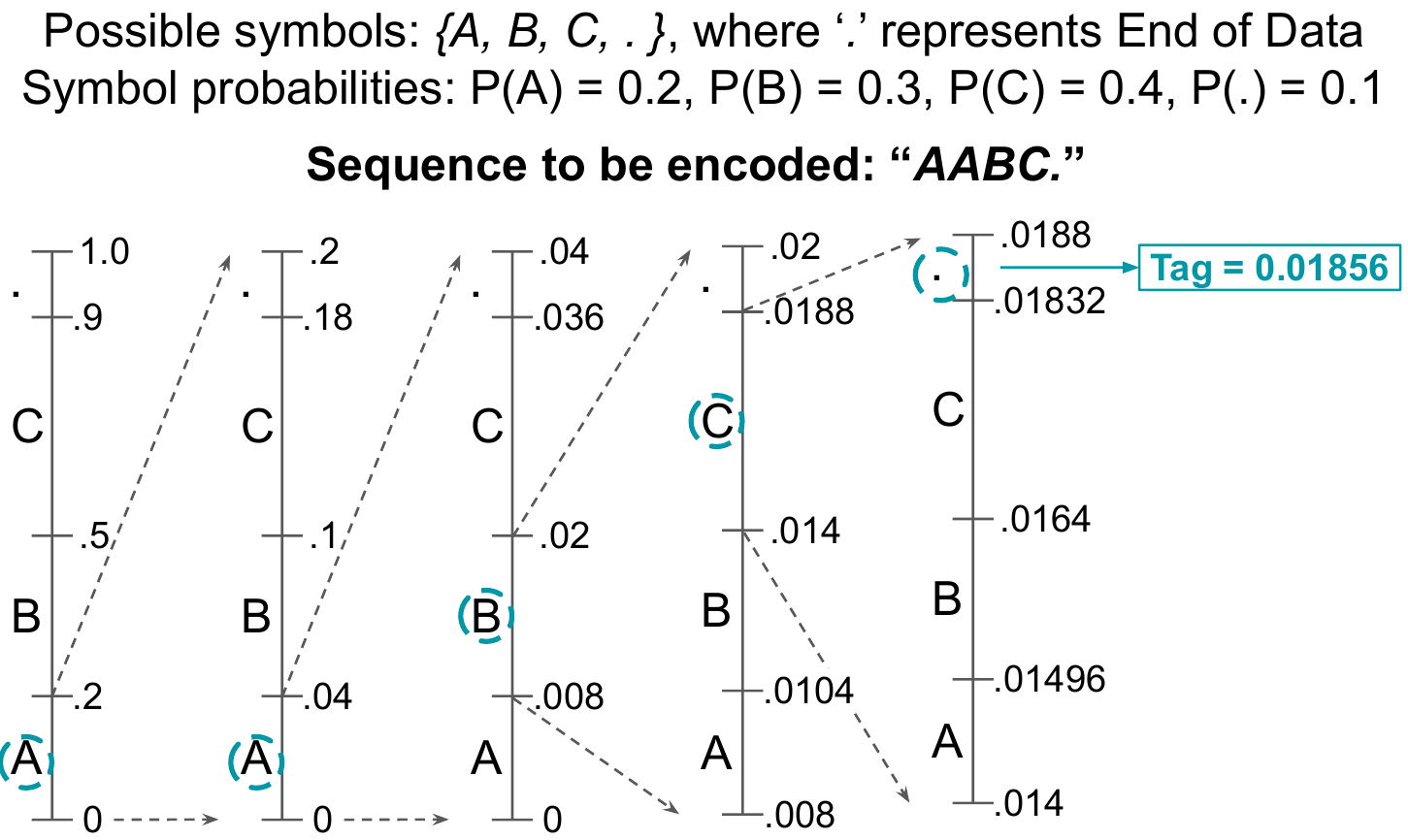}
  \caption{Given a finite set with all possible symbols and a probability distribution, the sequence "AABC" is encoded into a single decimal value. The process starts by dividing the range [0, 1) proportionally to the input distribution. Then, the process picks the segment that corresponds to the first symbol from the original sequence for further division. This process is repeated recursively until all symbols have been encoded.}
  \label{fig:arithmetic_coding_compression}
\vspace{-0.5cm}
\end{figure}

Our method leverages arithmetic coding (AC) \cite{10.1145/214762.214771} to compress the sequences of traffic values. This is a lossless method that compresses a stream of symbols (e.g., text characters) into a single number between [0, 1). To do this, AC assigns less bits to frequent symbols and more bits to less frequent symbols. In contrast to other popular compression methods such as Huffman coding~\cite{4051119}, AC achieves better compression ratios and it can work in an online fashion. In addition, the AC compression algorithm works with probability distributions, making it a good fit with ML technologies.

Figure~\ref{fig:arithmetic_coding_compression} shows the procedure to code a short text sequence using AC. Initially, the AC takes as input the set of possible symbols and a probability distribution. For simplicity, in this example the distribution remains static but a predictive model can be used to update the distribution after coding each symbol. Initially, the range [0, 1) is divided into segments proportionally to the symbol probability distribution. Then, the AC selects the segment [0, 0.2) corresponding to the first input symbol 'A' from the text sequence. Afterwards, this segment is divided into segments following the same proportions of the probability distribution. This process is repeated recursively for each symbol until the End-of-Data symbol is met. Finally, a decimal value from within the End-of-Data segment is picked as the tag for the entire text sequence.

\begin{figure}[!t]
  \centering
  \includegraphics[width=0.67\linewidth]{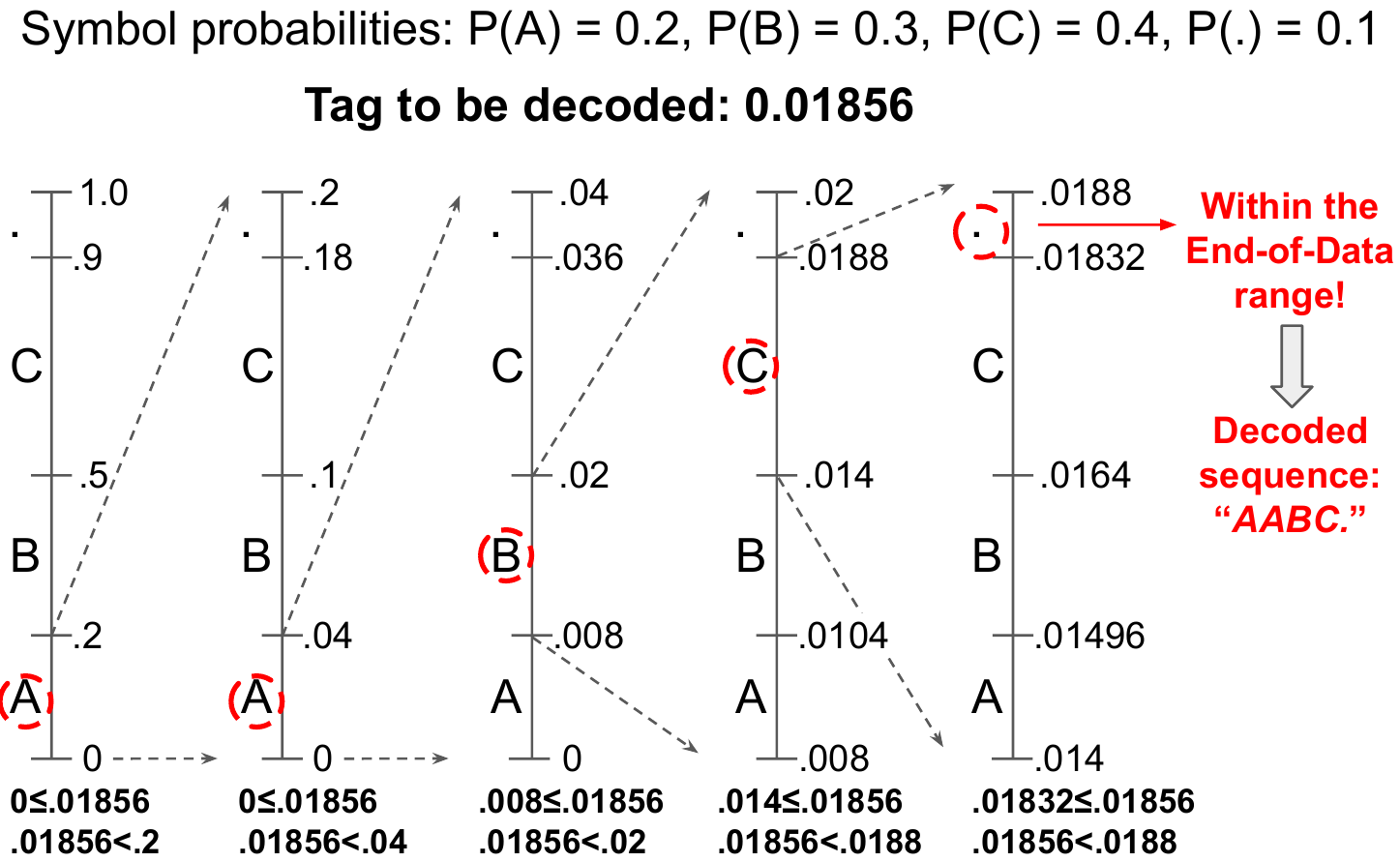}
  \caption{Given a tag, a set of symbols and a probability distribution, the decoding process starts by dividing the range [0, 1) proportionally to the distribution. The segment that contains the tag is selected and its corresponding symbol is decoded as the first symbol from the original text sequence. Then, the segment is divided proportionally, starting a recursive process that finishes with the End-of-Data symbol.}
  \label{fig:arithmetic_coding_decompression}
 \vspace{-0.3cm}
\end{figure}

The decoding part follows a symmetric procedure to the coding part. To recover the original text sequence, the algorithm takes as input the tag, the set of possible symbols and the probability distribution. Similarly, the range [0, 1) is divided into segments proportionally to the probability distribution and the segment that includes the codeword is selected. The symbol that corresponds to the segment represents the first symbol from the original text sequence. Then, the process starts again, decoding the original sequence one symbol at a time. This is a recursive process that finishes when the End-of-Data symbol is met. Figure~\ref{fig:arithmetic_coding_decompression} shows an example of decoding the tag and recovering the original text message.

The compression performance of the AC is defined by the quality of the probability distribution. Consider a scenario where there is a predictive model to dynamically compute the probability distribution for each symbol in a sequence. As an example, consider the AC at time bin \textit{t} and we want to compress the value at \textit{t+1}. Then, the AC can use a predictive model that takes as input the past \textit{k} symbols and predicts the probability distribution for the next symbol at \textit{t+1}. If the model is accurate, AC will assign less bits to encode the symbol, resulting in close-to-optimal compression performance. On the other hand, if the model is not accurate, the probabilities will not correspond to the real symbol, which results in poor compression or it can even increase the final file size. In this paper we want to leverage ML to build an accurate predictor to compress the sequences of network traffic measurements.

\begin{figure*}[!t]
  \centering
  \includegraphics[width=0.75\linewidth]{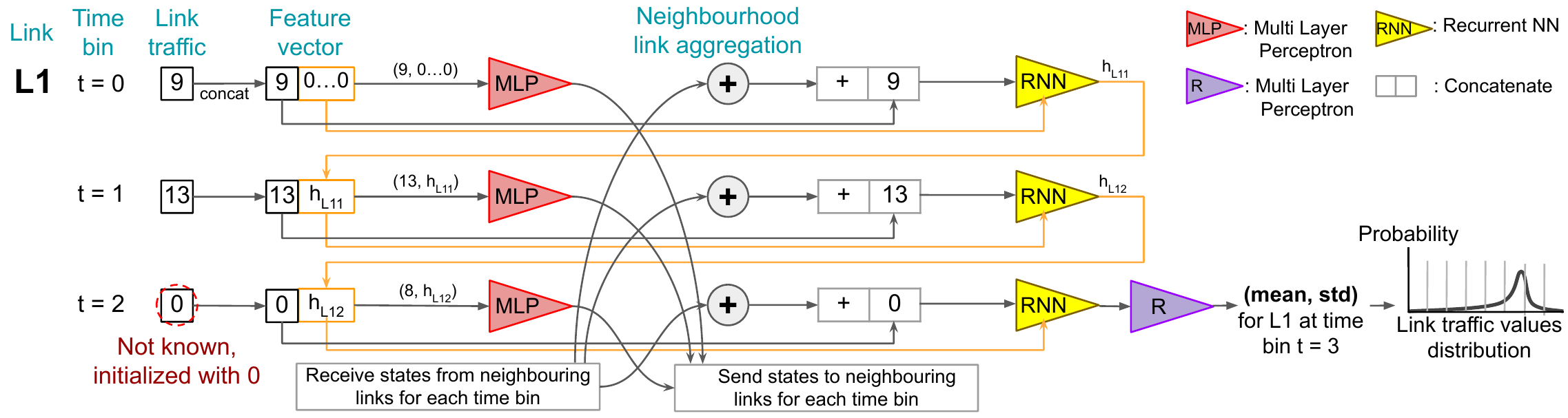}
  \caption{Initially, the per-link feature vectors are initialized at \textit{t=0} for all links with the corresponding traffic values and padded with zeros. These vectors are then processed by the MLP, resulting in a new hidden state vector. For each link in the topology, the hidden states of the neighboring links are aggregated (e.g., using a sum) and concatenated with the actual link hidden state. The resulting hidden state is processed by the RNN, which outputs a final hidden state for the current time bin. This state is then used to initialize the same link feature vector in time bin \textit{t=1}. The same process is repeated and it finishes after iterating over all bins within the time window. Finally, a different MLP (denoted R) takes the resulting hidden state from the last time bin and outputs the mean and standard deviation of a probability distribution. This distribution is used by the AC to code the actual traffic values of link \textit{L1} at time bin t=2.}
  \label{fig:space_time_mp}
 \vspace{-0.4cm}
\end{figure*}

\subsection{Notation and problem statement}
\label{subsec:notation}
Formally, link-level traffic measurements are represented as a matrix $\bm X\in \mathbb{N}^{w\times l}$, where $w$ represents the sliding window for $l$ links.
Each traffic measurement is a random vector $\bm x_t\in\mathbb{N}^{l}$.
For the arithmetic encoder we need a one step forecast distribution $p(\bm x_{t}| \bm x_{<t})$ to capture temporal dependence.
As the arithmetic encoder operates on streams of symbols, we further partition the distribution with a chain rule to capture spatial dependence:
\begin{equation}
    p(\bm x_{t}|\bm x_{<t}) = \prod_l p(x_l|\bm x_{t,<l},\bm x_{<t}).
 \label{fig:spatial_equation}
\end{equation}
Here we assume the stationary model $p(x_l|\bm x_{t,<l},\bm x_{<t})$ and mask-out the unknown traffic values.
Note that there is no natural order for the auto-regressive model, however, the only requirement is that the order must be the same during compression and decompression.

\begin{figure}[!b]
  \centering
  \includegraphics[width=0.8\linewidth]{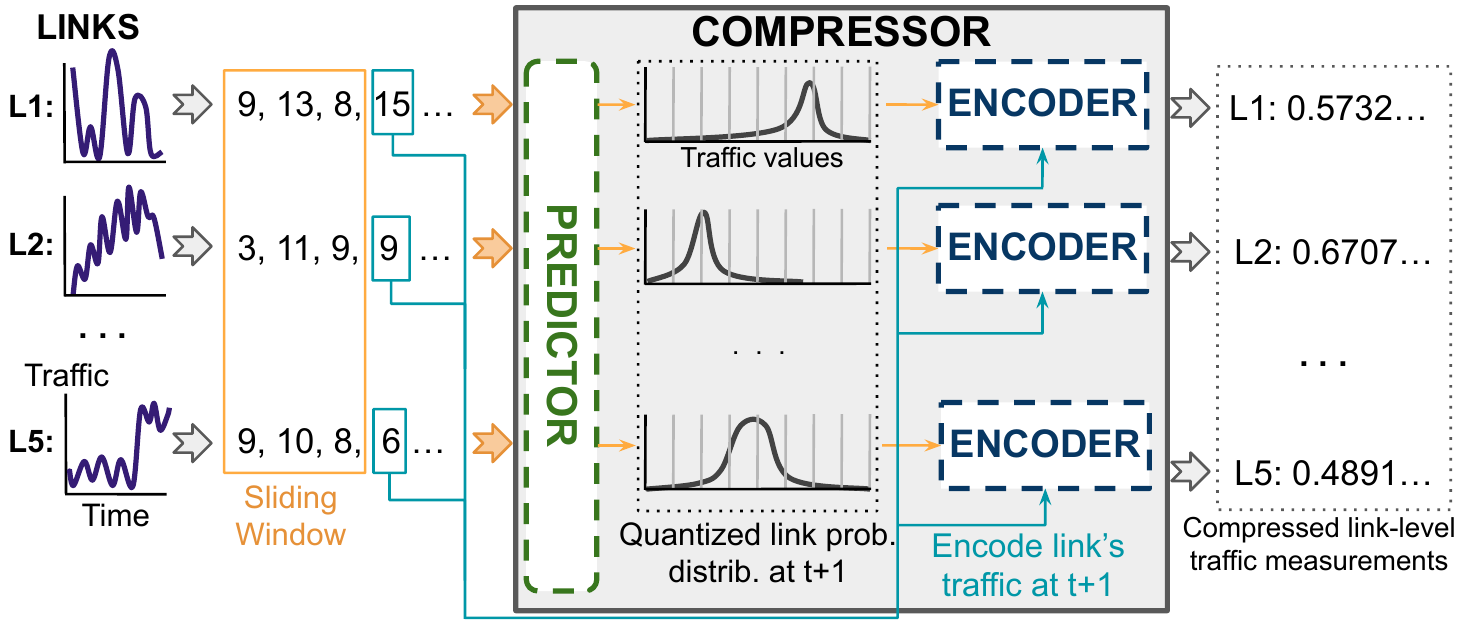}
  \caption{Our method takes as input the traffic values from within the sliding window and outputs a tag per link, which represents the per-link compressed sequence. In the network-wide scenario, the network topology is also part of the input data. Suppose the window finishes at time bin \textit{t}, the predictor computes the probability distribution of the traffic values at \textit{t+1}. These are used to encode the real values from \textit{t+1}. Afterwards, the sliding window is shifted by one time bin and the process starts again until the end of the sequence..}
  \label{fig:archi_overview}
\end{figure}

\section{Design}
\label{sec:design}

In this section we present a method for network traffic compression based on Neural Networks (NN). This method compresses link-level traffic measurements that evolve during time. These measurements are performed periodically and aggregated in time bins. For example, the bins can be of 5 minutes, indicating the traffic that passed through a link during this period of time.

We consider two link-level compression scenarios. In the first one, we want to compress link-level traffic measurements from a single link (e.g., access link). This is a common practice in small or medium size networks where internal traffic is smaller and not considered to be of interest in many cases (e.g., enterprise, campus networks). With a single link, we can only exploit temporal correlations as no other links are considered. The second scenario represents a more general use-case, where we want to simultaneously compress the traffic from multiple links of a network topology (i.e., network-wide compression~\cite{10.1145/1015467.1015492}). This situation will be more common in large networks, such as those of Internet Service Providers, which can have a global view of the network topology. In this case, apart from the temporal correlations of the first scenario, the routing and the topology will also introduce spatial correlations that we can exploit for compression purposes.

Our compression method takes as input the link-level traffic values and outputs a floating point number for each link. This value corresponds to the tag of the AC and it represents the entire sequence compressed (see Section~\ref{subsec:arithm_coding}). In the network-wide scenario (e.g., ISP), our proposed method takes as input the network topology in addition to the traffic values. The proposed compression method uses a sliding window to iterate over all traffic measurements. In other words, it processes the traffic values within the sliding window to output a probability distribution. This distribution is used to actually encode the traffic measurements immediately after the sliding window. The process is repeated until the window iterated over all values.

The proposed compression method is composed of two main blocks: the predictor and the encoder/decoder. The predictor leverages a NN model to predict the probability distribution in the next time bin after the sliding window (see Section~\ref{subsec:predictor}). The encoder/decoder is in charge of actually compressing the sequences of traffic values (see Section~\ref{subsec:encoder_decoder}). The better are the predictions of the NN model, the better is the compression ratio of our method. Figure~\ref{fig:archi_overview} shows an overview of the compressor module, with its inputs and outputs.

\subsection{Predictor}
\label{subsec:predictor}

The predictor is implemented using a Recurrent Neural Network (RNN) in its simplest form. This implementation is used when compressing link-level traffic measurements of a single link. Specifically, the RNN processes the link-level sequence of traffic values and afterwards a Multi-Layer Perceptron (MLP) takes the resulting hidden states and outputs the parameters of a probability distribution (e.g., Normal, Laplace). The probability distribution is then used by the AC to code the real link measurements. Notice that the RNN architecture only exploits temporal correlations. 

In the network-wide scenario, we implement the predictor using a Spatio-Temporal Graph Neural Network (ST-GNN)~\cite{10.5555/3304222.3304273}. Inspired by the message passing neural network~\cite{gilmer2017neural}, the proposed ST-GNN uses a message passing step for each time bin to exploit the spatial and temporal correlations. This step consists of exchanging information between neighboring links and it is necessary to propagate the link-level information across the topology. The ST-GNN takes as input the traffic measurements and the network topology and outputs a per-link probability distribution. The ST-GNN enables to exploit both spatial and temporal correlations naturally present in network traffic traces (see Section~\ref{subsec:exploiting_temp_spati}).

\begin{figure*}[!t]
  \centering
  \includegraphics[width=.8\linewidth]{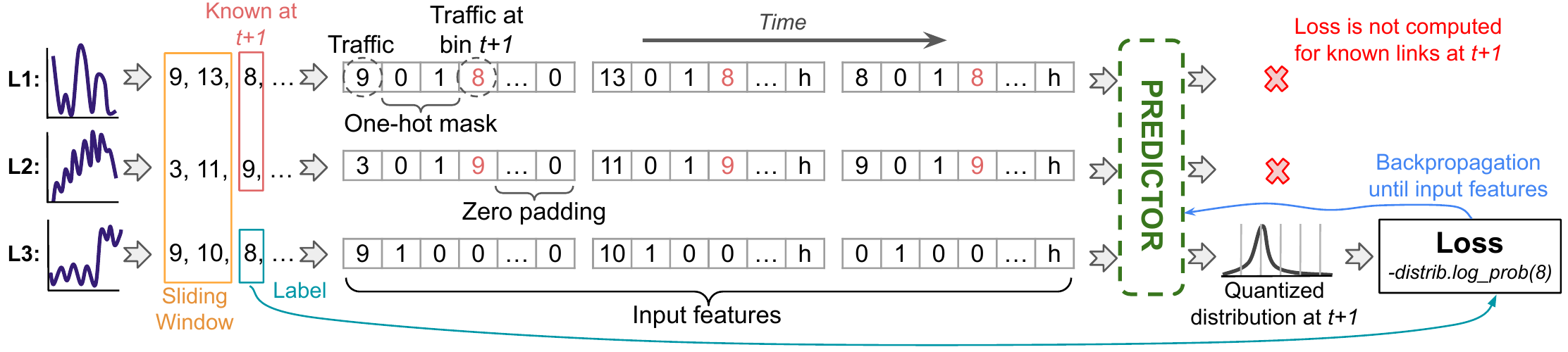}
  \caption{Graphical example showing how link features are initialized for each time bin. Consider a time window of size 2 that finishes at time bin \textit{t}. From time bin \textit{t+1}, we already compressed 2 traffic values from link \textit{L1} and \textit{L2} and we want to compress the value in \textit{L3}. The feature vectors assigned to each link for each time bin are initialized with the known values and the mask. This information is processed by the predictor, which outputs a quantized probability distribution for the missing link value at \textit{t+1}. The loss is computed for the same link and back propagated all the way to the link input features.}
  \label{fig:link_features}
  \vspace{-0.3cm}
\end{figure*}

Figure~\ref{fig:space_time_mp} shows an overview of the internals of the ST-GNN architecture. For simplicity, we only show the steps of predicting the probability distribution in a single link \textit{L1} using a time window of size 2. Initially, at time bin \textit{t=0} the links' feature vector are initialized with the traffic values and padded with 0. Then, the feature vector is processed by a MLP and the output vector is sent to all the neighboring links. At the same time, the current link \textit{L1} receives the hidden states from the neighboring links, aggregates them using a sum and concatenates the actual link traffic value. The resulting hidden state is then processed by the RNN, which outputs a final hidden state for the present time bin. This hidden state is used in the next time bin \textit{t=1} to initialize the feature vector. The process is repeated for all time bins within the sliding window. In the last bin, a different MLP (denoted R in the figure) is used to process the final hidden states and to output the mean and standard deviation of the traffic value probability distribution (e.g., Normal, Laplace). This probability distribution is used by the AC to compress the traffic value of link \textit{L1} at time bin \textit{t=2}.

\subsection{Encoder/Decoder}
\label{subsec:encoder_decoder}

The encoder/decoder is responsible for effectively compressing/decompressing the actual sequences of traffic values. Specifically, it takes the link-level probability distributions from the predictor and compresses/decompresses the sequences of traffic values. We implement the encoder/decoder using arithmetic coding (AC)~\cite{10.1145/214762.214771}. The AC compresses each link-level traffic sequence into a single floating point number (i.e., one decimal number per link). When decoding, the method works symmetrically to the encoder (see Section~\ref{subsec:arithm_coding}). The more accurate the NN-based predictor, the higher the compression ratios as less bits will be used for compressing the traffic sequences.

\subsection{Mask}
\label{subsec:mask}

The ST-GNN model uses a mask to exploit the spatial correlations between links, enabling the model to learn the conditional distribution $p(x_l|\bm x_{t,<l},\bm x_{<t})$ from Equation~\ref{fig:spatial_equation}. Specifically, the mask is used to gradually incorporate the already compressed/decompressed link traffic values of a time bin into the prediction. By masking the known link traffic values, our model learns to predict the conditional probability distribution. As an example, consider a topology with 2 links and a time window of size 2. This means we know all the link traffic values for time bins \textit{t=0} and \textit{t=1}. Then, the ST-GNN uses the known traffic values to predict the probability distributions for both links at \textit{t=2}. From all the distributions, a single one is picked and the corresponding link traffic value is compressed using AC. The link is then marked as known for bin \textit{t=2} using the mask and the ST-GNN uses the updated link features to predict the probability distribution for the missing link. This prediction is conditional to the known traffic value, and thus exploiting the spatial correlation between links.

During training, the mask of unknown links is created randomly. This means that for each sliding window we associate a random mask over the links to indicate whose link traffic values are known. In the compression/decompression phase, the mask starts by marking all the traffic values as unknown. Then, the predictor compresses/decompresses the traffic values in order, incorporating them into the prediction by changing the mask. Figure~\ref{fig:link_features} illustrates how the link-level features are initialized and how the loss is computed for a single link. In particular, for each link and time bin, the input link features are the traffic measurements and the mask.

\subsection{Compression}
\label{subsec:compression}

\begin{figure}[!b]
  \centering
  \includegraphics[width=0.8\linewidth]{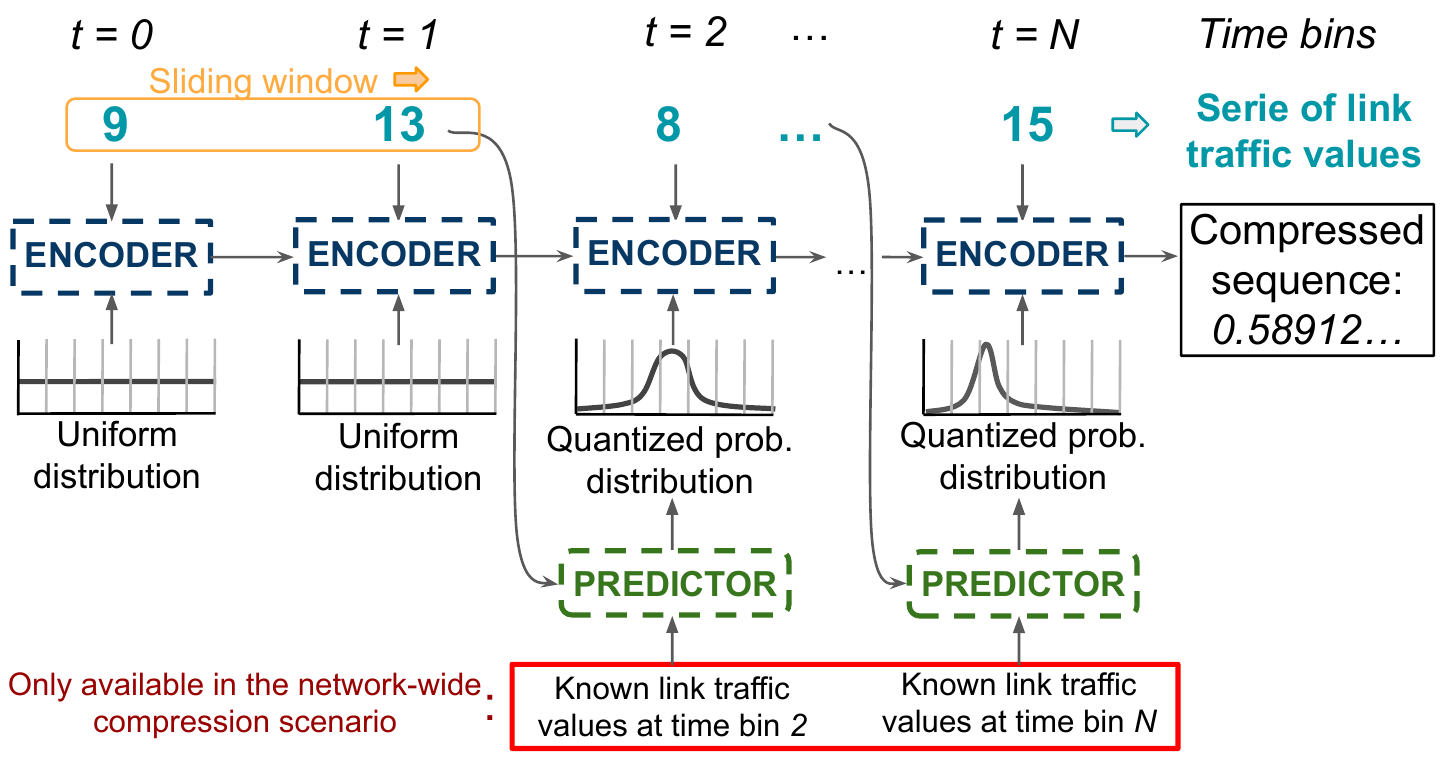}
  \caption{Overview of the compression process for a single link. The predictor processes the information from the sliding window. In the network-wide scenario, the predictor incorporates information from the links whose traffic is known at \textit{t+1} to predict the conditional probability distribution $p(x_l|\bm x_{t,<l},\bm x_{<t})$. The output is a single decimal value that encodes the link's traffic sequence.}
  \label{fig:encoder_framework}
\end{figure}

\begin{algorithm}[!t]
\caption{Compression algorithm}\label{compression_algo}
\begin{algorithmic}[1]
\State \textbf{Inputs:} $seq\_windows$ \Comment{Sequence of ordered windows}
\State $first\_w\gets True$
\ForEach{$ w \in seq\_windows$}\label{line:loop1} 
\If{$first\_w$}
\State $prob\_dist\gets uniform\_dist$\label{line:unif}

\ForEach{$ bin \in range(len(w)-1)$}\label{line:loop2}
\ForEach{$ link \in range(num\_links)$}\label{line:loop3}
\State $encode(w[link,bin], prob\_dist)$
\EndFor
\EndFor \label{line:end_unif}
\State $first\_w, bin\gets False, len(w)$\label{line:len_bin}
\EndIf

\ForEach{$ l \in range(num\_links)$}\label{line:loop4}
\State $prob\_dist\gets model\_inference(w)$ \label{line:model_inference}
\State $prob\_dist\gets quantize(prob\_dist)$\label{line:quantization}
\If{$scenario==network-wide$}
\State $link \gets select\_link(prob\_dist)$
\label{line:select_link}
\State $encode(w[link,bin], prob\_dist)$ \label{line:encode_link1}
\State $w\gets update\_link\_features(w,link)$  \label{line:update_features}
\Else
\State $link \gets l$
\State $encode(w[link,bin], prob\_dist)$  \label{line:encode_link2}
\EndIf

\EndFor
\EndFor
\end{algorithmic}
\end{algorithm}

Our proposal uses the link-level traffic values from the sliding window to predict the probability distributions, including the masked links from the next time bin. In particular, if the time window finishes at time bin \textit{t}, it uses the previous \textit{k} traffic values until \textit{t} to predict the probability distributions for time bin \textit{t+1}. These distributions are then used by the AC to code the actual values from time bin \textit{t+1}. Figure~\ref{fig:encoder_framework} shows an overview of the compression process for a single link. In the network-wide compression scenario, the predictor is implemented with a ST-GNN that takes as additional input features the neighboring link's hidden states. When compressing a single link, the predictor is implemented with a RNN that has only information from the past traffic values independently for each link.

Algorithm~\ref{compression_algo} shows in detail how the compression procedure works. To simplify the pseudocode, we compress the traffic values in the last time bin from a sliding window. The algorithm takes as input the ordered sequence of time windows and starts iterating over them (line~\ref{line:loop1}). The first traffic values from the first time window are compressed using uniform probabilities (line~\ref{line:unif} to line~\ref{line:end_unif}). Then, the algorithm encodes the traffic values from the last position of the time window (line~\ref{line:len_bin}). To do this, a loop iterates over each link, compressing one value at a time (line~\ref{line:loop4}). For each link, the algorithm leverages the NN-based model to compute the quantized probability distributions (lines~\ref{line:model_inference} and~\ref{line:quantization}). The model can be implemented with a RNN or a ST-GNN, depending on the compression scenario. In the network-wide scenario, the algorithm uses a heuristic to determine the link order (line~\ref{line:select_link}), incorporating them into the prediction. After encoding the selected link, the link-level features and the mask are updated in line~\ref{line:update_features}. The process starts again and is repeated until all links have been encoded.

The heuristic to select the link in the network-wide scenario is based on an increasing order of standard deviation (line~\ref{line:select_link}). We experimentally observed this heuristic helps the ST-GNN decrease the uncertainty in the predictions. In other words, leaving the links where the GNN model is more uncertain to the end helps the GNN make better predictions. This is because the ST-GNN will have more links with known traffic values, reducing the model's uncertainty over the links with higher standard deviation.

The compression process results in a tag for each link, encoding the link's traffic sequence. This tag is stored in a single file on disk. When decompressing, the tag is used by the decoder to recover the original traffic sequence without losing information. Notice that the compression process is performed in a streaming fashion, differing from traditional methods like GZIP that are static. In other words, our method can compress the traffic values as they come, without the need of storing the measurements in a buffer before compressing.

\subsection{Decompression}

\begin{figure}[!b]
  \centering
  \includegraphics[width=0.8\linewidth]{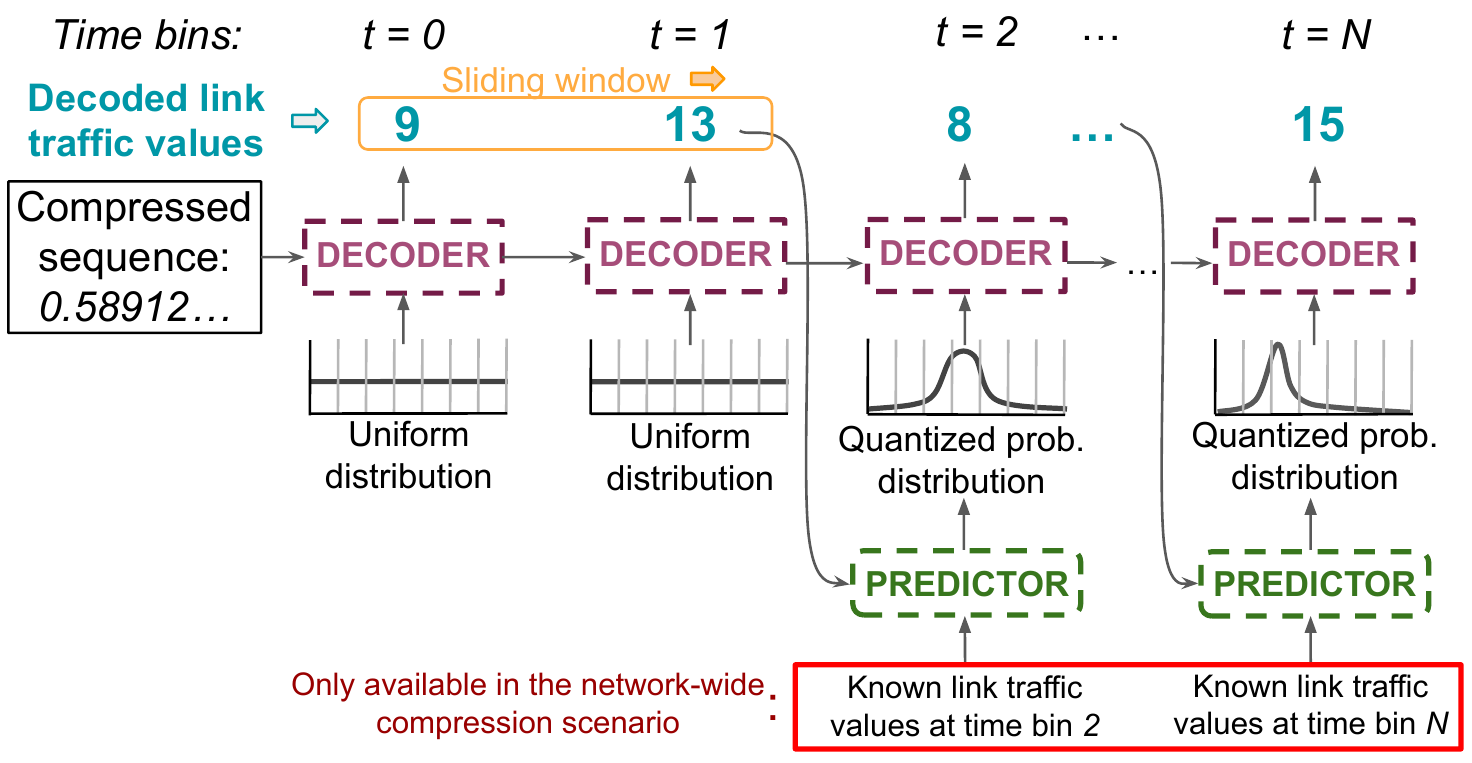}
  \caption{Overview of the decompression process for a single link. The process is similar to the compression, but now the arithmetic coding uses the decoder to recover the original sequence of symbols. Notice that for each time bin the predictor receives the same input information as in the compression phase.}
  \label{fig:decoder_framework}
\end{figure}

For decompression, the model reads the tag from the compressed files and uses the same NN-based model to recover the original link sequences bin by bin. The first elements within the first sliding window are decompressed using uniform probability distributions. Then, when the entire sliding window is decompressed, the predictor uses the recovered values to compute the probability distributions for the links. Similarly, in the network-wide compression scenario the algorithm leverages the same heuristic to select the order in which to decode the traffic values for each time bin. Figure~\ref{fig:decoder_framework} shows a general overview of the decompression phase. Notice that for each time bin, the predictor receives the same input information as in the compression phase. The decoder also receives the same information but in this case its operations are inverted to decode (see Section~\ref{subsec:arithm_coding}). The algorithm to decompress is the same as Algorithm~\ref{compression_algo} but replacing the coding operations from \textit{encode(·)} by their inverse decoding operations.

\section{Experimental Results}

\subsection{Methodology}
\label{subsec:methodology}

We evaluated the compression performance of our method with respect to GZIP, the \textit{de facto} compression method of network traffic traces. In the first experiment, we generated synthetic datasets with different intensities of spatial and temporal correlations on the NSFNet topology \cite{Hei:04} with 42 directional links. We synthetically created signals with different correlations (see Section~\ref{subsec:synth_data_gen}) and extracted 1,000 samples using a window size of 5 time bins, including the labels. In other words, the NN-based models leverage the link-level traffic values of 4 time bins to compress the values within the 5th bin. During training, we created 50 different random masks for each time window. We experimentally observed that the higher the number of different masks, the better is the accuracy of the ST-GNN model, but paying the cost of increasing training time. This is because increasing the number of different masks enriches the training process as there is more data to train the NN on.

In the second experiment, we evaluated the compression capabilities of the NN-based models on three real-world datasets. The first two datasets are the Abilene and Geant datasets from \cite{SNDlib10}. The Abilene dataset corresponds to a topology with 30 directional links and a total of 41,741 samples after data cleaning and using a time window of size 5. This dataset contains the link-level traffic measurements (in bytes) during 6 months in intervals of 5 minutes. The Geant dataset corresponds to a topology of 72 directional links and a total of 6,063 samples after data cleaning for the same window size. The third dataset is more recent and it was obtained from in-house link-level traffic measurements in a campus network. The dataset contains 12 months of per-link network traffic measurements in intervals of 5 minutes (from December 2020 until December of 2021).  The topology contains 16 directional links and a total of 102,799 samples with window size of 5. 

All the experiments were performed on \textit{off-the-shelf} hardware. In particular, we used a single machine with an AMD Ryzen 9 5950X 16-Core Processor with one GeForce GTX 1080 Ti GPU for training the models. We trained all NN-based models using 70\% of the samples for training and 30\% for evaluation. For each time bin, we created 40, 40 and 20 unique random masks for the Abilene, Geant and Campus network datasets. The loss function used was the negative log likelihood of a Laplace distribution. To work with a finite set of probabilities, we quantized the Laplace distribution. After training, we chose the model with lowest evaluation error and we compressed the entire datasets.

\subsection{Implementation}

We implemented the ST-GNN and the RNN using Tensorflow 2.8~\cite{tensorflow2015-whitepaper}. The RNN was implemented using the Gated Recurrent Unit architecture \cite{chung2014empirical}. The scripts to pre-process the datasets were written in Python 3.8 and we used the NetworkX~\cite{hagberg2008exploring} and NumPy~\cite{harris2020array} libraries for graph-related operations. We leveraged an open-source implementation of the arithmetic coding for Python\cite{nayuki} to implement the encoder. In the synthetic experiment, we used the Statsmodels Python library\cite{seabold2010statsmodels} to implement the auto-regressive model.

\subsection{Synthetic data generation}
\label{subsec:synth_data_gen}

In our compression scenario, we assumed the network topology is an input to the model (only in a network-wide scenario). The only remaining variables that have an impact on the link-level traffic measurements are the source-destination flows. There is one flow for each pair of source-destination nodes within a network topology. All flows follow the shortest path routing policy to reach the destination node. Consequently, each link will be traversed by a subset of all flows.

We assigned for each flow a periodic signal that originated from a \textit{sine} wave. This wave is scaled by 40 to obtain values in the order of hundreds when aggregated on each link and is shifted to contain only positive values. In addition, we add random noise to the signal, we randomly shift its phase to start at different values and we randomly change the periodicity. If all flows originate from the same signal, the links are highly correlated in space as their values will increase/decrease proportionally for each time bin. In our experiment, we consider 4 degrees of spatial correlation: 0\%, 30\%, 60\% and 100\%, indicating the percentage of flows that have the same signal characteristics.

\begin{figure}[!t]
  \centering
  \includegraphics[width=0.99\linewidth]{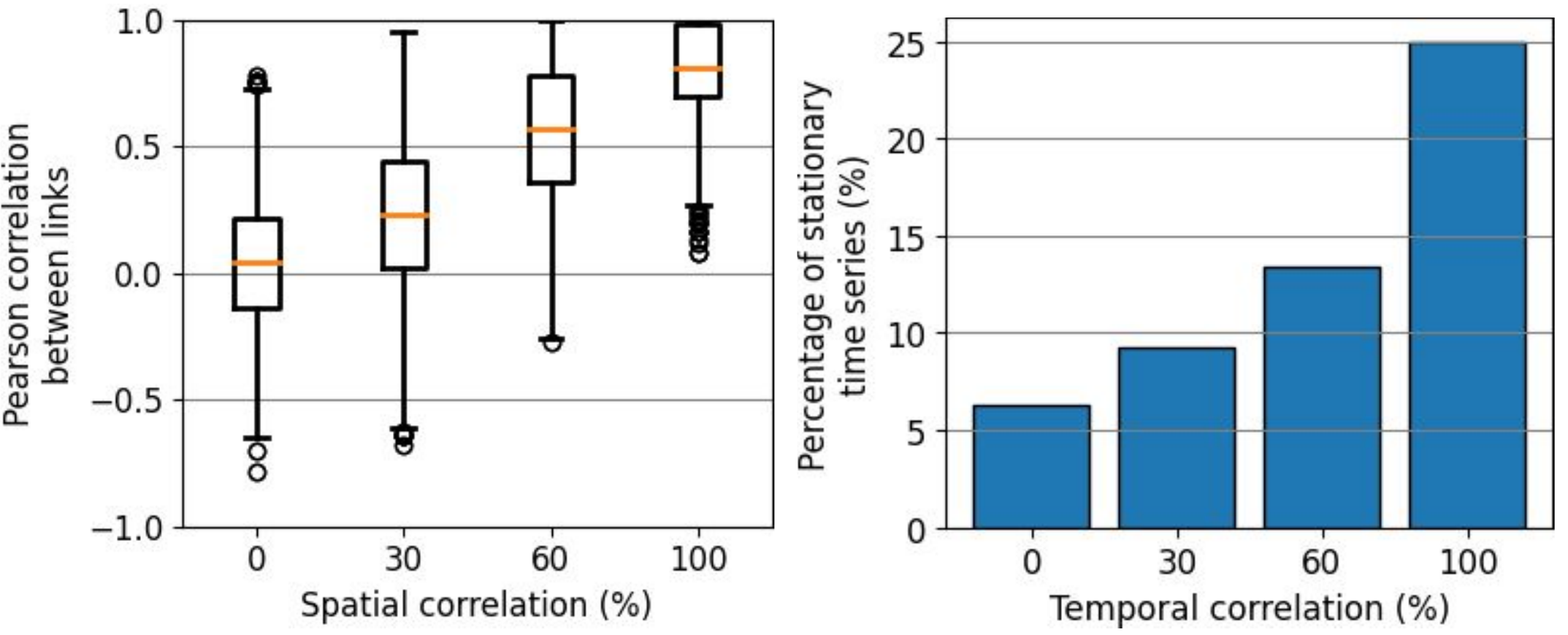}
  \caption{Boxplots of the Pearson correlation in the synthetic datasets (left). The higher is the spatial correlation, the higher are the pearson correlation coefficients between links. On the right, we show the higher is the temporal correlation, the  higher is  the percentage of link-level stationary time series in the synthetic datasets.}
  \label{fig:temp_spati_corr_evol}
  \vspace{-0.5cm}
\end{figure}

\begin{figure*}[!t]
  \centering
  \includegraphics[width=0.7\linewidth]{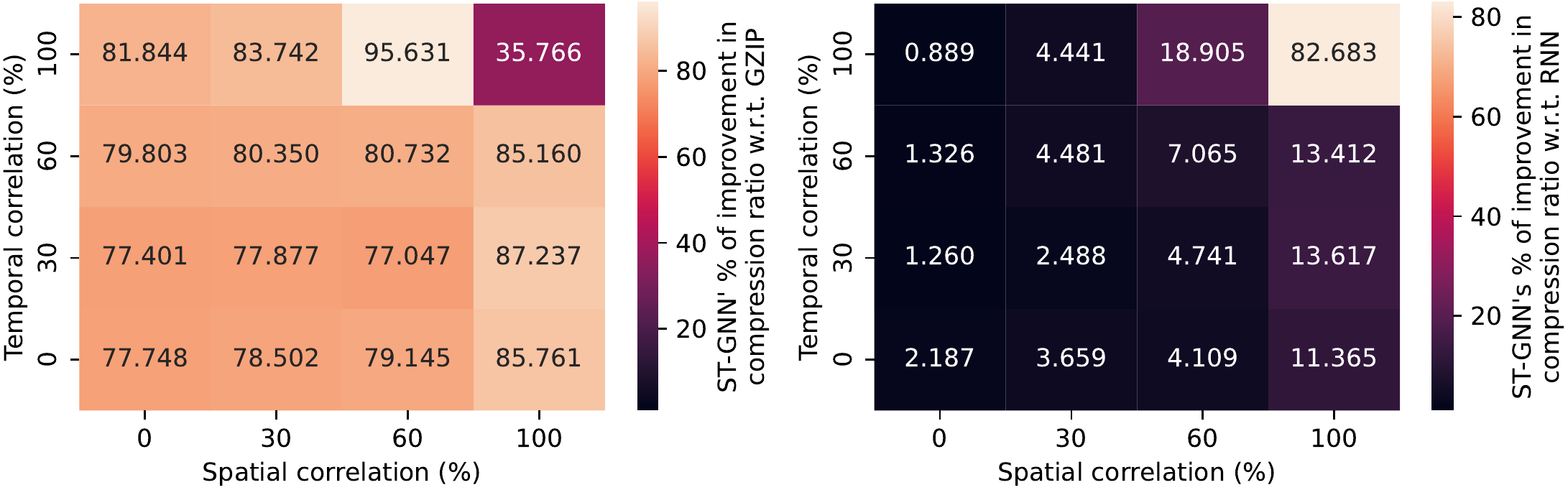}
  \vspace{-0.2cm}
  \caption{Compression ratio improvement for the ST-GNN model with respect to GZIP (left) and RNN (right). Notice that in the scenario with higher spatial and temporal correlations there are a few traffic values that are highly repeated in the dataset, which GZIP's underlying algorithm exploits effectively.}
  \label{fig:heatmap}
  \vspace{-0.3cm}
\end{figure*}

Temporal correlations are present in stationary time series with periodical patterns. The \textit{sinusoidal} signal naturally meets these requirements, resulting in a time series with high temporal correlation. To decrease the temporal correlation, we added additional noise to the flow signal from auto-regressive (AR) models. To control the intensity of the temporal correlation in the experiment we controlled the percentage of flows that are added with noise. Specifically, we consider 4 degrees of temporal correlations: 0\%, 30\%, 60\% and 100\%, indicating the percentage of flows that conserve the original signal.

In Figure~\ref{fig:temp_spati_corr_evol} (left) we show how the spatial correlation increases in our synthetic datasets. Specifically, we grouped all the datasets by the intensity of spatial correlation. Then, we computed the Pearson correlation for each pair of links following Equation~\ref{equation:pearson_correlation}. The results indicate that the higher the intensity of the spatial correlation (x-axis), the higher the Pearson correlation (y-axis). Figure~\ref{fig:temp_spati_corr_evol} (right) shows how the temporal correlation present in the synthetic data increases with the temporal correlation coefficient (x-axis). Similarly, we grouped all datasets by temporal intensity. Then, we performed the Augmented Dickey-Fuller (ADF) test~\cite{dickey1979distribution} for each link-level traffic sequence, counting the number of paths that are stationary (i.e., the mean and variance of the time series do not change in time). In particular, if the ADF test was returning a \textit{p-value} smaller or equal than 0.05, we rejected the null hypothesis (H0), considering the time series to be stationary.

\subsection{Evaluation on synthetic data}

In total, there are 16 experiments that correspond to all possible combinations of spatial and temporal correlation intensities. For each of these experiments, we consider both network-wide and independent link-level compression scenarios (see Section~\ref{sec:design}). In total, we trained 32 models from which 16 of them were based on the ST-GNN model and the other 16 on RNNs.

Figure~\ref{fig:heatmap} (left) shows the percentages of compression ratio improvement for the ST-GNN with respect to the GZIP baseline in the network-wide scenario. The results indicate that the ST-GNN outperforms GZIP by a large margin in all experiments. Notice that the scenario with maximum spatial and temporal correlation contains a small number of link-level traffic values that are repeated frequently. GZIP uses Huffman coding~\cite{4051119} as the underlying algorithm, which can effectively exploit the repeated traffic values. This explains why the compression ratio improvement is the lowest for this particular scenario. In addition, the figure indicates the expected performance of the ST-GNN when evaluated in real-world scenarios. In particular, the intensity of the temporal and spatial correlations of a real-world dataset could point to the expected performance with respect to GZIP.

To showcase the capabilities of our method to exploit spatial and temporal correlations simultaneously, we compare it with the RNN-based model. Recall that the RNN compresses one link only. Therefore, we apply the RNN model for each link in the topology, exploiting temporal correlations solely. For the sake of fairness, we maintained the same hidden state sizes in both ST-GNN and RNN models. 

Figure~\ref{fig:heatmap} (right) shows the performance improvement of the ST-GNN models with respect to the RNN-based models. The ST-GNN model outperforms the RNN in all correlation scenarios, but it has outstanding performance in scenarios with high spatial correlation. The results indicate that our model has the flexibility to exploit both spatial and temporal correlations. Notice that in the case of 0\% of spatial correlation and maximum temporal correlation, the improvement of the GNN model is $\approx$1\%, indicating that they perform similarly when there is high temporal correlation. This is expected as the ST-GNN model also incorporates a RNN (see Figure~\ref{fig:space_time_mp}).

\begin{figure}[!b]
  \centering
  \includegraphics[width=0.7\linewidth]{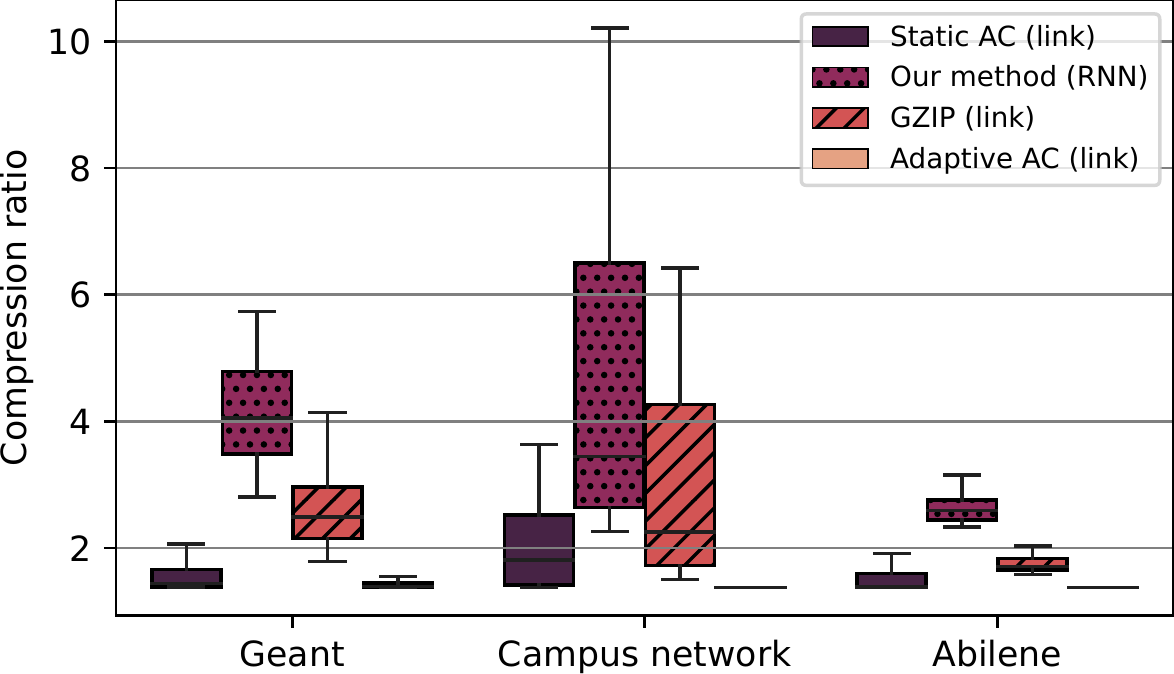}
  \caption{Compression ratios for the single-link scenario.}
  \label{fig:link_level_compression}
\end{figure}

\subsection{Compressing real-world data}
\label{subsec:comp_real_world}

In this experiment we evaluated the compression performance of our method on real-world link-level traffic measurements. To do this, we compressed three real-world datasets (see Section~\ref{subsec:methodology}). We compared the compression performance against three baselines: Static AC, Adaptive AC and GZIP. The Static AC and Adaptive AC baselines are similar to our method but the probability distribution is computed without using ML. In particular, Static AC iterates over the entire dataset and computes the probability distribution for the link-level traffic measurements. Then, it compresses/decompresses the entire dataset using the same static distribution for each AC coding step. The Adaptive AC baseline computes the new distribution using the values within the sliding window. These baselines are intended to show the benefit of using ML to implement the predictor model. Finally, we apply GZIP to compress the entire dataset.

Figure~\ref{fig:link_level_compression} shows the compression ratios for the three real-world datasets in the link-level scenario. In particular, each baseline was applied to compress each link individually and we compared the resulting compressed links with their original file size (i.e., one file per link). The results indicate a remarkable performance of our compression method for all datasets, outperforming GZIP by a large margin. In addition, the figure indicates a clear advantage of using an adaptive ML-based predictor to exploit temporal correlations present within the sliding window.

\begin{figure}[!t]
  \centering
  \includegraphics[width=0.7\linewidth]{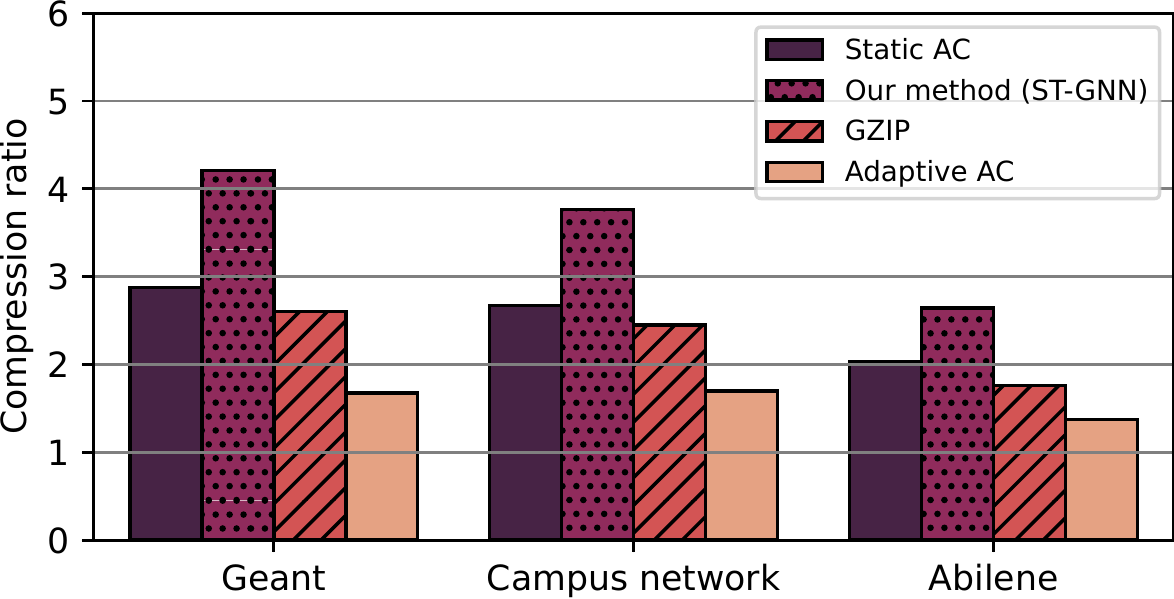}
  \caption{Compression ratios for the network-wide scenarios. Notice that in this experiment we are compressing the entire dataset.}
  \label{fig:bar_plot_comp_ratios}
 \vspace{-0.5cm}
\end{figure}

Figure~\ref{fig:bar_plot_comp_ratios} shows the experimental results of compressing the same datasets in the network-wide scenario. Recall that our compression method is implemented using the ST-GNN, which leverages the traffic values from all links to dynamically compute the probability distributions. In this scenario, Static AC computes the probability distribution using the entire dataset (i.e., including all links) and Adaptive AC updates the distribution by including the values from all links within the sliding window.

The experimental results indicate that our proposed method achieves the highest compression ratios in all three datasets. Particularly, it outperforms Static AC by $\approx$47\%, $\approx$41\% and $\approx$29\% for the Geant, Campus network and Abilene datasets respectively. In addition, the performance improvement with respect to GZIP is of $\approx$62\%, $\approx$53\% and $\approx$50\% for the same datasets respectively. These results showcase the benefit of leveraging ML to exploit spatial and temporal correlations for traffic compression.

\subsection{Cost}

In this section we discuss the cost of using our method for online compression. Specifically, our method compresses the traffic measurements in a streaming fashion (see Section~\ref{sec:design}). This means that it can compress the link traffic measurements as they come from the network monitoring platform. Conversely, GZIP needs to wait to have the entire dataset to apply the compression algorithm. Alternatively, GZIP could compress all links at once on each time bin independently. We did this and the experimental results indicate that GZIP achieves a compression ratio of $\approx$0.94, $\approx$0.4 and $\approx$0.54 for the Geant, Campus network and Abilene datasets. In other words, the compressed data occupies more space than the original data,
contrasting with the results from Figure~\ref{fig:bar_plot_comp_ratios}.

We computed the average cost of compressing one time bin using the ST-GNN and RNN models to evaluate the deployment to real-world online traffic compression. In addition, we computed the size of storing the model's weights into a file. Table~\ref{table:cost} shows the average cost of compressing one time bin for all real-world datasets, indicating that our method is capable of online compression. In other words, when it receives the aggregated traffic of, for example, the last 5 minutes, our method can effectively compress the values in the order of seconds. Finally, Table~\ref{table:cost} also shows how much memory it is required to store the trained weights of the neural network. The results indicate that the model is lightweight and it achieves high compression ratios with an expendable model size.

\vspace{-0.2cm}

\begin{table}[!t]
\centering
\begin{tblr}{
  colspec = {crrrr},
  cell{1}{2,4} = {c=2}{c}, 
}
\hline[2pt]
  Dataset &   Mean Cost (s) &      &  Model size (Kbytes) &      \\
\hline[1pt]
       &    ST-GNN &  RNN &  ST-GNN & RNN \\
\cline{2-5}
    Geant  &  3.20 & 0.47 & 575 & 489 \\
    Campus network  & 0.22  & 0.07 & 236 & 202 \\
    Abilene  &  1.22 &  0.29 & 332 & 285 \\
\hline[2pt]
\end{tblr}
\vspace{0.1cm}
\caption{Model size and mean cost (in seconds) to compress one time bin.}
\label{table:cost}
\vspace{-0.7cm}
\end{table}

\section{Related Work}

The most popular existing method for network traffic compression is GZIP. However, the networking community has investigated different approaches for compressing network traffic. The work of \cite{aiello2005sparse} proposes a lossy method to approximate the real network traffic by capturing the most relevant traffic features. In \cite{beirami2015packet} they propose to exploit the traffic redundancies at the packet level to reduce the transmitted traffic. In \cite{mehboob2006high} they propose an architecture to implement the LZ77 compression algorithm \cite{ziv1977universal} on a FPGA. The work of \cite{fusco2010net} describes a solution for on-the-fly storage, indexing and querying of network flow data. A more recent work leverages the P4 language~\cite{bosshart2014p4} and generalized deduplication to implement a solution that operates at line-speed.

The compression method presented in this paper has similarities with the problem of traffic prediction. The works of \cite{vinayakumar2017applying, ramakrishnan2018network} propose the use of RNNs to predict the network traffic. A more recent work proposes to use simulated annealing and an Autoregressive Integrated Moving Average model \cite{yang2021network} to predict the network traffic. In~\cite{andreoletti2019network} they use a graph-based ML algorithm to predict the link-level traffic loads in backbone networks. The work from~\cite{gao2020predicting} they leverage inter-flow correlations and intra-flow dependencies to predict the traffic matrix using RNNs. Finally, the work from \cite{li2020deep} proposes to use a spatio-temporal convolutional neural network with attention mechanisms to predict wireless traffic. 

Despite the similarities with traffic prediction, it is important to remark that traffic compression has some particularities. First, our compression method considers only the probability distribution for each link, instead of the exact traffic values. This is due to the requirements of the arithmetic coding part. Second, we only work with the values from the next time bin, whereas in traffic prediction the work horizon is typically larger (e.g., predict the traffic for the next couple of hours). Finally, when decompressing information we do not have access to the first elements of the time-window, forcing the use of simple methods that do not depend on the compressed information (e.g., uniform probability).

\section{Conclusion}

Existing methods for network traffic compression are generic, resulting in low compression ratios. This limitation becomes even more critical when compressing traffic in an online scenario. In our work, we proposed the use of ML and arithmetic coding to compress link-level traffic measurements. Specifically, we presented a method that exploits the spatial and temporal correlations intrinsic in the traffic measurements. The experimental results show that it can effectively compress real-world traffic traces, with an improvement of $\geq$50\% in compression ratio for real-world datasets with respect to GZIP.

\section{Acknowledgment}

This publication is part of the Spanish I+D+i project TRAINER-A (ref.~PID2020-118011GB-C21), funded by MCIN/ AEI/10.13039/501100011033. This work is also partially funded by the Catalan Institution for Research and Advanced Studies (ICREA) and the Secretariat for Universities and Research of the Ministry of Business and Knowledge of the Government of Catalonia and the European Social Fund. This work was also supported by the Polish Ministry of Science and Higher Education with the subvention funds of the Faculty of Computer Science, Electronics and Telecommunications of AGH University and by the PL-Grid Infrastructure. 

\bibliography{references}
\bibliographystyle{unsrt}

\end{document}